# Are spiral disks really opaque?

C. Xu[1] and V. Buat[2] [3]

[1] Max-Planck-Institut für Kernphysik, Postfach 103980, D69117 Heidelberg, Germany
[2] Laboratoire d'Astronomie spatiale du CNRS, BP 8, 13376 Marseille Cedex 12, France
[3] Laboratoire des intéractions photons-matière, Faculté des Sciences et Techniques de St Jérome, 13397 Marseille Cedex 13, France

**Abstract.** We compare the ultra-violet, optical, and far-infrared emission for a sample of 135 spiral galaxies in order to address the widely debated problem concerning the opacity of spiral disks. We find that the re-radiation of the dust, estimated from the far-infrared emission, is on average only $31\pm1$ percent of the bolometric luminosity of a spiral galaxy, indicating that less than one third of the stellar radiation is absorbed and then re-radiated by dust in a spiral disk. Applying a radiation transfer model which assumes a 'Sandwich' configuration for the spiral disk, and fully takes into account the effect of scattering, we find for our sample a median of the face-on blue (4400Å) optical depth $\tau_B = 0.49$ and the mean $<\tau_B> = 0.60 \pm 0.04$, indicating that most spiral galaxies in our sample are *not* opaque for blue light ($\tau_B < 1$).

## 1. Introduction

The question whether the disks of spiral galaxies, which consist of stars, gas and dust, are transparent (optically thin) or opaque (optically thick) in optical wavelength range, is currently a hotly debated one. It is equivalent to the question "Are the spirals intrinsically very different from what we see from the Earth?" Our understanding of the nature of these astronomical systems, which contain most of the visible material of the universe, depends fundamentally on the ultimate answer to this question. The classical approach to the problem is based on the statistical studies of the dependence of the optical surface brightness $\mu$ on the view angle i (i = 0 for seen face on), under the hypothesis that for a transparent disk $\mu$ depends sensitively on i, while for an opaque disk it does not (Holmberg 1958). Applying this method to a sample of more than 9000 disk galaxies, the largest sample to date, Valentijn (1990) claims that most spiral disks are opaque ($\tau_B > 1$), and therefore are intrinsically much brighter than observed. However, this approach is rather indirect and is influenced sensitively by biases in the sample selection (Disney 1992). Consequently, conflicting results have been obtained by different authors (Holmberg 1958; de Vaucouleurs 1959; Tully 1972; Davis et al. 1989; Valentijn 1990; Burnstein et al. 1991; Boselli & Gavazzi 1994).

Modern astronomy has opened the entire domain of the electromagnetic waves, from radio ($\lambda \gtrsim 1$ cm) to $\gamma$–ray ($\lambda \lesssim 0.1$Å). All the energy radiated by a spiral disk can, in principle, be recorded. The problem of the optical thickness can be addressed directly by comparing the escaped emission of stars, the major sources of emission of galactic disks, with the re-radiated emission of dust which causes the extinction. If the disks are indeed opaque, the latter should be much larger than the former, and vice versa. Stars emit predominantly in ultra-violet (UV: 500–3650Å), optical (3650–9000Å), and near-infrared (NIR: 9000Å – 3$\mu$m); dust emits in mid-infrared (MIR: 5–40$\mu$m) and far-infrared (FIR: 40–1000$\mu$m). In this paper, we compare these two emissions for a sample of spiral galaxies and find that most spiral disks in our sample are *not* opaque ($\tau_B < 1$).

## 2. Dust emission to bolometric emission ratio

Our sample is selected from nearby spiral galaxies detected in two recent UV experiments: FAUST (Bowyer et al. 1993; Deharveng et al. 1994) is a far-UV telescope (central wavelength 1650Å) flown in 1992 on board of the space shuttle Atlantis; and SCAP (Donas et al. 1987) is a balloon-boarded UV experiment (central wavelength 2030Å) carried out in Marseille during the early 1980s. Galaxies are included if passing the following two criteria: (1) detected by IRAS in the 60 and 100 $\mu$m bands, (2) morphological types from Sa to Sm. The sample contains 135 galaxies. The monochromatic UV fluxes are taken at 2030 Å, the FAUST data are translated to this wavelength according to the calibrations of Deharveng et al (1994) (a systematic difference of 29% is found between the FAUST and SCAP fluxes). The IRAS 60$\mu$m and 100$\mu$m fluxes are taken from the IRAS data base. The monochromatic 4400Å fluxes $f_{4400}$ are taken from the Third Reference Catalog of Bright Galaxies (de Vaucouleurs et al. 1990, hereafter RC3). Both the UV and the blue fluxes are corrected for the foreground Galactic extinction using the blue extinction tabulated in the RC3 and the extinction law of Savage and Mathis (1979).

In order to estimate the integrated UV flux $f_{UV}$ (912 – 3650Å) from the monochromatic flux $f_{2030}$, the average broad band UV spectra (in the range of 1400 – 3650 Å) reported by Pence (1976) for Sa–Sab galaxies, and by Coleman et al. (1980) for Sb–Sbc, Sc–Scd, and Sd–Sm galaxies are extrapolated to the range of 912 – 3650Å, and then are adopted for the corresponding types. The optical-NIR fluxes $f_{op-nir}$ are generally estimated from the blue fluxes using the average broad band optical-NIR spectra for different Hubble types estimated from the unweighted mean of the empirical spectra collected by Yoshii & Takahara (1988), except for 85 galaxies whose B – V colors are taken from RC3, and for 52 galaxies whose U – B colors are taken from the same source. In a first time we neglect the ionizing UV radiation (< 912Å) which is predominantly absorbed by dust. Finally, we estimate the total escaped radiation of stars from a disk galaxy by $f_{star} = f_{UV} + f_{op-nir}$.

The dust emission is estimated from IRAS fluxes $f_{60\mu}$ and $f_{100\mu}$, namely the radiation in the wavelength range of 40 — 120$\mu$m. An important issue is the radiation of the cold dust grains ($\sim$ 15K) at wavelengths > 120$\mu$m. There is evidence from IRAS studies and mm/sub-mm observations that such cold grains dominate the total *mass* of dust (Devereux & Young 1990; Kwan & Xie 1992; Bothun & Rogers 1992; Chini & Krügel 1993). However, because of the strong dependence of FIR luminosity on grain temperature ($L \propto T^{4+n}$, n=1 $\sim$ 2), the contribution of cold grains to the total *emission* of the dust is not very significant (Devereux & Young 1990; Devereux & Young 1994). For example, recently Guélin et al. (1993) find that the 1.3mm continuum flux of NGC 891 is about a factor of 8 more than a simple extrapolation of the IRAS fluxes suggesting the existence of significant amount of cold dust, but the luminosity of this cold dust ($L_{cd}$) is only 1/3 of that of the 'warm dust' ($L_{wd}$). Nonetheless, large uncertainties are present when only IRAS data and 1mm data are compared (Devereux & Young 1990; Valentijn 1991), and different conclusions have been reached by different authors; for instance Chini et al. (1986), fitting the IRAS data and their 1.3mm data using a two-component model, found that the cold dust emission (18K) dominates the warm dust (53K) emission. The best way to estimate the cold dust radiation beyond 120$\mu$m is perhaps to compare the IRAS data and the sub-mm data (mostly in 300–800$\mu$m) which are much closer to the peak of the dust emission (likely in 100–200$\mu$m). Recently, using available sub-mm data (Chini et al. 1986; Stark et al. 1989; Eales et al. 1989), Kwan & Xie (1992) estimated the total FIR luminosity (40–1000$\mu$m) for 11 late-type galaxies. Adding NGC 891 (Guélin et al. 1993) and the Galaxy, the FIR–sub-mm spectrum compiled by Mazzei et al. (1992, see also Chini & Krügel 1993), we find a mean of L(40–1000$\mu$m)/L(40–120$\mu$m) = 1.39 $\pm$ 0.21 for these 13 galaxies.

the 25$\mu$m band (18–30$\mu$m). Hence, again for the 13 galaxies mentioned above, we estimate the MIR fluxes (8–40$\mu$m) using the formula

$$f_{mir} = f_{12\mu} \times \frac{c\,\delta\lambda_1}{\lambda_1^2} + f_{25\mu} \times \frac{c\,\delta\lambda_2}{\lambda_2^2} \qquad (1)$$

where $f_{12\mu}$ and $f_{25\mu}$ are in Jy, c is the speed of light, $\lambda_1 = 12\mu$m and $\delta\lambda_1 = 7\mu$m, and $\lambda_2 = 25\mu$m and $\delta\lambda_2 = 25\mu$m. The value of $\delta\lambda_2$ is larger than the real bandpass of IRAS 25$\mu$m channel (11.15$\mu$m) in order to cover the wavelength range 15–40 $\mu$m. The total $f_{dust}/f_{fir}$ ratio, $f_{fir}$ being the integrated FIR fluxes in the wavelength range 40–120$\mu$m estimated from $f_{60\mu}$ and $f_{100\mu}$ (Helou et al. 1988), is estimated as

$$\frac{f_{dust}}{f_{fir}} = \frac{f_{mir}}{f_{fir}} + \frac{f(40-1000\mu m)}{f_{fir}} \qquad (2)$$

In Fig.1 we plot the $f_{dust}/f_{fir}$ ratio versus $\log(f_{60\mu}/f_{100\mu})$ for the 13 galaxies with sub-mm data. The ratio spans from 1.4 to 2.2, with a mean 1.79$\pm$0.25. A strong linear anti-correlation between $f_{dust}/f_{fir}$ and $\log(f_{60\mu}/f_{100\mu})$ is found (linear correlation coefficient -0.76). This is easy to understand because the warmer the FIR color ratio ($f_{60\mu}/f_{100\mu}$), the more contribution from the warm component to the total dust emission (Xu & De Zotti 1989; Kwan & Xie 1992). The dotted line is the linear regression

$$f_{dust}/f_{fir} = 1.45\,(\pm 0.08) - 1.06\,(\pm 0.24) \times \log(f_{60\mu}/f_{100\mu}) \qquad (3)$$

We shall adopt this empirical relation for calculations of $f_{dust}$ from $f_{fir}$ and $f_{60\mu}/f_{100\mu}$. It should be noticed that this relation is derived from a small sample of galaxies, and there might be some systematic uncertainties with their sub-mm data (Chini & Krügel 1992). Our results might be affected by these uncertainties.

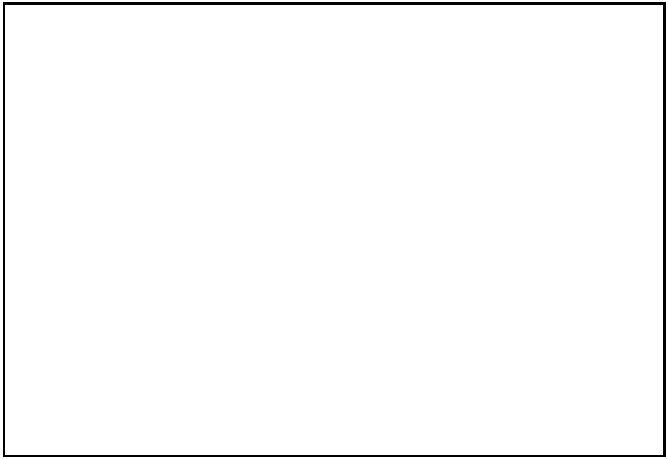

**Fig. 1.** Diagram of $f_{dust}/f_{fir}$ ratio versus $\log(f_{60\mu}/f_{100\mu})$ for: 1) 11 galaxies with sub-mm data taken from Kwan & Xie (crosses); 2) NGC 891 (open square); 3) the Galaxy (solid square).

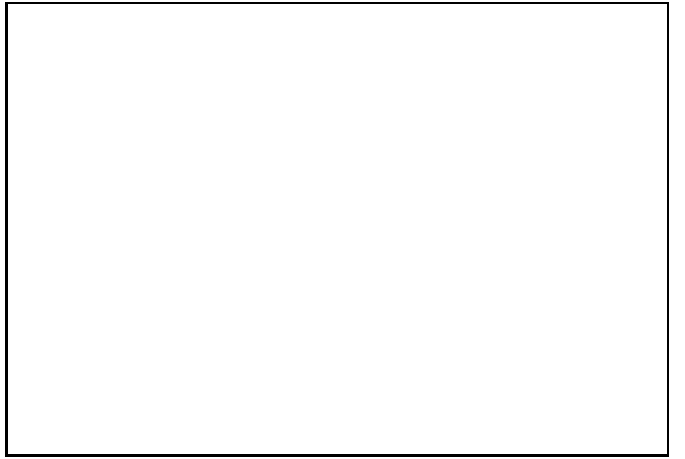

**Fig. 2.** Histogram of dust–to–bolometric emission ratio. The bolometric flux $f_{bol} = f_{star} + f_{dust}$.

There is also 'hot dust' consisting of small-grains/large-molecules heated transiently by single UV or optical photons to T > a few hundred K. It emits at wavelengths shorter than 40$\mu$m (MIR), and may account for 20–30% of the total dust emission (Puget & Léger 1989; Désert et al. 1990). Most of this emission is included in the IRAS 12$\mu$m band (8–15$\mu$m) and

Figure 2 is a histogram of the $f_{dust}/f_{bol}$ distribution for our sample, where the bolometric flux $f_{bol} = f_{star} + f_{dust}$. The sample mean is 0.31$\pm$0.01, indicating that on average the dust absorbs less than one third of the stellar radiation. This is consistent with the result of Cox and Mezger (1989), who find that the dust emission is only one third to one fourth of the bolometric luminosity of the Galaxy.

## 3. A detailed radiation transfer model

In the literature, the opacity of disks is usually studied in terms of $\tau_B$, the optical depth for blue (4400Å) radiation. In the following we estimate $\tau_B$ for the galaxies in our sample using a detailed model for radiation transfer and dust heating.

We consider three contributions to the heating of dust: 1) dust heating due to the ionizing radiation ($\lambda < 912$Å), 2) due to the non-ionizing UV radiation ($912 < \lambda < 3650$Å) and 3) due to the optical radiation ($3650$Å $< \lambda < 9000$Å). We assume that the ionizing radiation mainly heats the dust around the massive stars with which the HII regions are associated. Subtracting this contribution from the total dust radiation, we estimate the optical depth of the disk by comparing the remaining (diffuse) dust emission, the non-ionizing UV and the optical radiation for a sample of galaxies using a radiation transfer model.

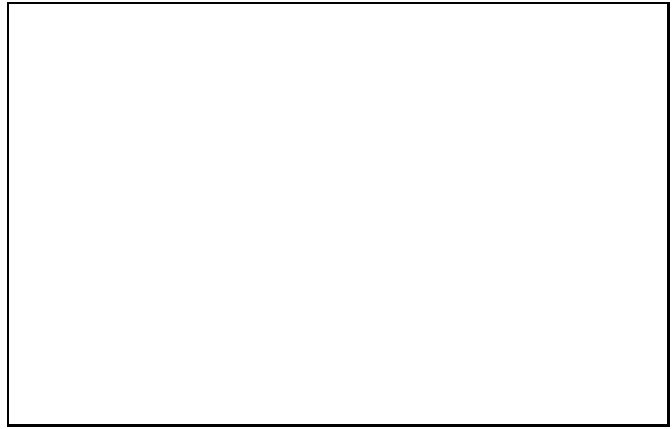

**Fig. 3.** Schematic illustration of definitions of several parameters used in the radiation transfer model. $\epsilon$ denotes the volume emissivity, $\kappa$ the volume absorptivity, $\mu' = \cos i'$, and $\tau_1 = \int_0^{l_1} \kappa\, dz$

The ionizing UV radiation ($\lambda < 912$Å) can be estimated from the H$\alpha$ emission. For 34 galaxies in the sample, the H$\alpha$ fluxes $f_{H\alpha}$ are available from Kennicutt and Kent (1983) and Romanishin (1990) (by order of preference). After correcting $f_{H\alpha}$ for 0.8 mag internal extinction (Kennicutt 1989), the ionizing flux is estimated as follows:

$$f_{Lyc} = N(Lyc) \times <E(Lyc)> = 33.9 \times f_{H\alpha} \qquad (4)$$

where $N(Lyc) = 7.19\ 10^{11}/0.75 \times f_{H\alpha}$ is the number flux (in units of cm$^{-2}$ sec$^{-1}$) of Lyman-continuum photons (Mezger 1978; Lequeux 1980), and $<E(Lyc)> = 2.13 \times E(Ly\alpha) = 3.49\ 10^{-11}$ erg is the average energy of Lyman-continuum photons (Mezger 1978). Assuming that 80 % of Lyman-continuum emission is absorbed by dust, directly or indirectly (via emission lines) and in or outside HII regions, the contribution from the ionizing UV radiation to $f_{dust}$ can be estimated:

$$f_{dust}^{lyc} = 0.8 \times f_{Lyc}\,. \qquad (5)$$

For 11 early spirals (Sa–Sbc) with $f_{H\alpha}$, we find that on average the ionizing UV contributes 14±2% of total $f_{fir}$. For the 23 later galaxies the contribution is 20±1%. We estimate the values of $f_{dust}^{lyc}$ for galaxies without H$\alpha$ data using these means. The rest of $f_{dust}$, $f_1 = f_{dust} - f_{dust}^{lyc}$, is due to non-ionizing UV and optical radiation.

The radiation transfer model for the non-ionizing UV and optical radiation of disk galaxies is illustrated in Fig.3. It is based upon the model of Xu & Helou (1994) which has been applied to the heating of the diffuse interstellar dust in M31. The model takes the effect of scattering fully into account in the sense that scattered light of any order has been calculated using an iteration procedure from lower order scattered light (van de Hulst & de Jong 1969). An infinite-plane-parallel geometry is adopted for the radiation transfer problem, and the radiation sources (stars) and dust are assumed to be smoothly distributed. We allow for different thicknesses of the star layer and of the dust layer ('Sandwich model'). The model calculates for a given wavelength, at a given point in the disk and in a given direction, the radiation intensity $I_\lambda(\tau_1, \mu')$ as a function of the radiation source function $S_{0,\lambda}$ and the optical thickness of the disk $\tau_\lambda$. From this we predict the ratio between the radiation absorbed by dust (dust-heating) and the radiation escaping the disk and eventually being observed, as a function of the optical depth of the disk and the view angle:

$$Q_\lambda(\tau_\lambda,\mu) = \frac{\int_0^{\tau_\lambda}\left[\int_{4\pi} I_\lambda(\tau_1,\mu')d\omega\right](1-a)d\tau_1/\mu}{I_\lambda(0,\mu) + E_1/\mu} \qquad (6)$$

where $I_\lambda(0,\mu) + E_1/\mu$ is the prediction for the light observed (see Fig.3 for the definition of $E_1$), $\mu = \cos i$ the inclination, $\tau_\lambda$ the face-on optical depth, $\omega$ the solid angle, and $a$ the albedo. The factor $(1-a)$ gives the ratio between the absorption cross section and the extinction cross section.

We split the dust heating into that due to the non-ionizing UV radiation (UV-heating) and that due to the optical radiation (optical-heating). The UV-heating is calculated as

$$\begin{aligned}f_{dust}^{UV} =& f_{2030} \times Q(\tau_{2030},\mu) \times \\ & \int_{912\text{Å}}^{3650\text{Å}} 10^{-0.4(m(\lambda)-m(2030\text{Å}))} \times \frac{Q(\tau_\lambda,\mu)}{Q(\tau_{2030},\mu)}\, d\lambda\end{aligned} \qquad (7)$$

where $f_{dust}^{UV}$ is the dust radiation intensity due to the UV heating; $f_{2030}$ the 2030 Å flux, and $\tau_{2030}$ the optical depth at 2030 Å; $(m(\lambda) - m(2030\text{Å}))$ is estimated from the adopted UV spectrum. Similarly, the optical heating is

$$\begin{aligned}f_{dust}^{op} =& f_{4400} \times Q(\tau_B,\mu) \times \\ & \int_{3650\text{Å}}^{9000\text{Å}} 10^{-0.4(m(\lambda)-m(4400\text{Å}))} \times \frac{Q(\tau_\lambda,\mu)}{Q(\tau_{4400},\mu)}\, d\lambda\end{aligned} \qquad (8)$$

The optical depth at different wavelengths (including $\tau_{2030}$) are related to $\tau_B$ through the adopted extinction curve.

The settings of the dust-heating program are
1) Sources of non-ionizing UV radiation (i.e. B stars) have the same scale height as dust ($E_1 = E_2 = 0$), while sources of optical radiation (i.e. old stars) are in a disk twice as thick as the dust layer so that half of these sources lie outside the dust layer, i.e. $E_1 = E_2 = 0.5 S_{0,\lambda} \times \tau_\lambda$.
2) The UV albedo = 0.18 (Hurwitz et al. 1991), and the scattering asymmetry factor $g = 0$ (isotropic scattering).
3) The optical albedo and phase function are taken from Mathis et al. (1983).
4) The UV extinction curves are taken from Pei (1992) for spiral galaxies.
5) The optical extinction curve is taken from Mathis et al. (1983).

For each galaxy in our sample, the face-on optical depth $\tau_B$ is estimated from the equation

$$f_{dust} = f_{dust}^{lyc} + [f_{dust}^{UV}(\tau_B) + f_{dust}^{op}(\tau_B)] \qquad (9)$$

where $f_{dust}^{lyc}$ is the dust emission due to ionizing UV flux. The results are plotted in Fig.4. The histogram of the $\tau_B$ distribution for all galaxies in the sample has a median value of 0.49, indicating that most of them are not opaque ($\tau_B < 1$). The mean value $<\tau_B> = 0.60 \pm 0.05$ is 'robust', i.e. the $3\sigma+$ points are not included in the calculation. The distribution is rather same stellar population may be responsible, through supernova explosion, for the acceleration of the cosmic-ray electrons radiating the nonthermal synchrotron radiation of these galaxies (Xu et al. 1994).

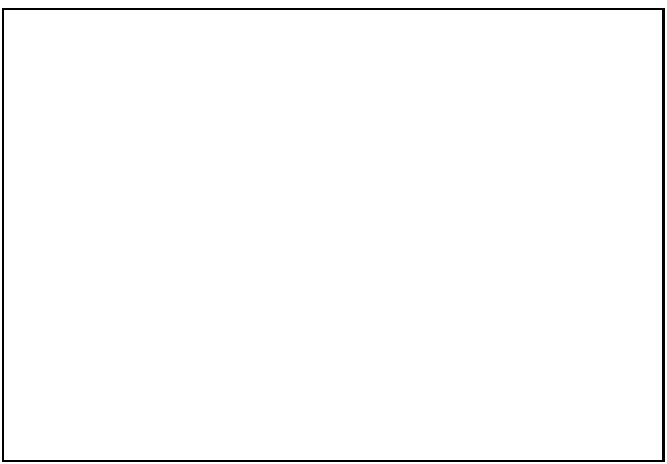

**Fig. 4.** Histograms of $\tau_B$ (face-on optical depth for radiation at 4400Å) distribution. Hatched: late spirals (Sc–Sm).

asymmetric with a sharp peak at $\tau_B \sim 0.2$ and a long tail at the large $\tau_B$ side. Therefore the mean is larger than the median. The histogram (hatched) of the $\tau_B$ distribution for late-type (Sc–Sm) spirals is similar to that of the total sample, with the median= 0.36 and the mean= $0.43 \pm 0.04$. Early type spirals (Sa–Sbc) are less transparent, their $\tau_B$ distribution is flatter than that of the late types with a median value of 0.86 and a mean equal to $0.79 \pm 0.08$. It is interesting to note that for 59 Sb–Sc galaxies in our sample, we find a median $\tau_B = 0.90$ and $<\tau_B> = 0.95 \pm 0.07$, slightly less than the result of Valentijn (1991) for the same types of galaxies ($\tau_B \simeq 1.3$).

Since our sample is basically UV selected, there might be a bias in the sense that less extinguished galaxies are more likely to be included. However, we find for our sample a mean $L_{fir}/L_B = 0.79 \pm 0.10$, where $L_{fir}$ is the integrated FIR luminosity in the wavelength range of 40–120$\mu m$, and $L_B$ the blue luminosity ($\nu L_\nu(4400Å)$). This is consistent with the FIR-to-blue luminosity ratio of UGC galaxies ($<L_{fir}/L_B> = 0.72$, Bothun et al. 1989), indicating that the bias, if existing, is not serious. A more detailed discussion of the extinction at UV wavelengths derived from this model is postponed to a subsequent paper (Buat & Xu, in preparation).

## 4. Discussion

Disney et al. (1989) exploit a similar method to the problem of the optical thickness of the spiral disks as we do in this letter. However, they compare the FIR emission of spiral galaxies only with the optical emission, neglecting the energy absorbed by dust from the UV radiation. They concluded that the data are consistent with the spiral disks being generally optically thick ($\tau_B > 1$). We reach a different conclusion because of including the contribution of the UV radiation to the dust heating. Indeed, we find that the UV radiation dominates the optical emission in dust heating: the fraction of dust heating due to non-ionizing UV (921Å $< \lambda <$ 3650Å) is 60±9%, and that due to optical radiation (3650Å $< \lambda <$ 9000Å) is only $21 \pm 8$%. The rest is due to the ionizing radiation ($\lambda < 912$Å). This is consistent with the result of Xu (1990) who found that the non-ionizing UV radiation can account for 57–76% of the radiation of the cool (diffuse) dust in spiral disks. The heating of dust by the non-ionizing UV radiation, radiated by intermediate massive stars ($\sim 5 M_\odot$) plays an important role in the remarkably tight and universal correlation between the FIR and radio continuum emission of spiral galaxies, because the